# Inverse Problems in Magnetohydrodynamics: Theoretical and Experimental Aspects

**Frank Stefani, Thomas Gundrum, Gunter Gerbeth, Uwe Günther, and Mingtian Xu**
*Forschungszentrum Rossendorf, P.O. Box 510119*
*D-01314 Dresden, Germany*
F.Stefani@fz-rossendorf.de

**ABSTRACT**

We consider inverse problems related to velocity reconstruction in electrically conducting fluids from externally measured magnetic fields. The underlying theory is presented in the framework of the integral equation approach to homogeneous dynamos in finite domains, which can be cast into a linear inverse problem in case that the magnetic Reynolds number of the flow is not too large. Some mathematical problems of the inversion, including the uniqueness problem in the sphere and a paradigmatic isospectrality problem for mean-field dynamos, are touched upon. For practical purposes, the inversion is carried out with the help of Tikhonov regularization using a quadratic functional of the velocity as penalty function. For the first time, we present results of an experiment in which the three-dimensional velocity field of a propeller driven flow in a liquid metal is reconstructed by a contactless inductive measuring technique.

**INTRODUCTION**

The determination of flow velocities in opaque fluids is a notorious problem in various industrial applications as, e.g., metallurgy and crystal growth. One important example is the continuous casting of steel for which one would like to have a rough picture of the actual flow pattern in order to react in case of irregularities. Another example is the Czochralski crystal growth technique for Silicon monocrystals for which some monitoring of the flow structure would be highly desirable. Needless to say that the usual optical flow measurement techniques, as Laser Doppler Anemometry (LDA) or Particle Image Velocimetry (PIV) are not applicable in those cases. Actually, there are a few measurement techniques for opaque fluids, including Ultrasonic Doppler Velocimetry (UDV), mechano-optical probes, and electric potential probes [1]. However, for hot and/or chemically aggressive materials it would be most desirable to avoid any contact with the melt.

This wish also applies to a particular sort of liquid sodium experiments which have been carried out since 1999 in order to investigate the homogeneous dynamo effect which is at the root of planetary, stellar, and galactic magnetism [2]. A contactless method to infer the sodium flow velocity from the externally measured magnetic fields would be of high value for the study of the saturated regime of those dynamos.

With view on these possible applications, the paper is intended to facilitate the use of a contactless inductive method for the velocity determination in electrically conducting melts, based on external magnetic field measurements. A clear shortcoming of the method, in comparison with traditional inductive flow determination based on electric potential measurements, is its restriction to fluid flows with not too small magnetic Reynolds numbers. Another shortcoming may be the fact that the method is not very accurate in the determination of local velocities. However, this point might be outweighed by the fact that the method enables online monitoring of transient flow fields in the entire fluid volume, at least with a time-resolution in the order of one second. This feature makes the method attractive for a number of technological and research applications.

The structure of the paper is as follows: We start with the theoretical set-up in the form of an integral equation system for the magnetic field and the electric potential that both are influenced (or even self-excited) by the velocity field. Then, the general form will be simplified for the case of small magnetic Reynolds numbers. The inverse task to reconstruct the velocity field from externally measured magnetic fields will be formulated.

Two short excursions will be put in, the first one dealing with the uniqueness problem for the general velocity reconstruction, the second one sketching an inverse spectral dynamo problem for a helical turbulence parameter.

Then we will switch over from pure theory to a laboratory experiment. Actually, this is a non-trivial step since the reliable determination of rather small induced magnetic fields on the background of large applied magnetic fields presents a serious problem for the measurement technique, at least, if one needs simultaneously magnetic field data at some dozens of sites at an affordable prize. In concrete, we present first results of an experiment with a propeller driven flow in a liquid metallic alloy that is exposed subsequently to two different external magnetic fields. For both applied magnetic fields, the induced magnetic fields are measured at 49 sites outside the fluid. From these two sets of data, the velocity field is derived by inversion. Although the reconstructed velocity structure is only rough, it reflects the essential characteristics of the flow quite realistically.

The paper closes with some conclusions and some ideas for future work.

## BASIC THEORY

When an electrically conducting medium, moving with the velocity $\mathbf{u}$, comes under the influence of an imposed steady magnetic field $\mathbf{B}_0$, an electromotive force (emf) $\mathbf{u} \times \mathbf{B}_0$ is induced that drives a current

$$\mathbf{j}_0 = \sigma(\mathbf{E}_0 + \mathbf{u} \times \mathbf{B}_0) \;, \qquad (1)$$

where $\sigma$ is the electrical conductivity of the fluid, and $\mathbf{E}_0$ is the electric field which results from the emf.

The current $\mathbf{j}_0$, in turn, produces an additional magnetic field $\mathbf{b}$, so that the total current $\mathbf{j}$ becomes dependent on the total magnetic field $\mathbf{B} = \mathbf{B}_0 + \mathbf{b}$

$$\mathbf{j} = \sigma(\mathbf{E} + \mathbf{u} \times \mathbf{B}) \;. \qquad (2)$$

The ratio of the induced magnetic field $\mathbf{b}$ to the imposed magnetic field $\mathbf{B}_0$ is governed by the so-called magnetic Reynolds number $Rm = \mu \sigma L U$, where $\mu$ is the magnetic permeability of the fluid, and $L$ and $U$ are typical length and velocity scales of the flow, respectively. This dimensionless number is, in the most relevant technical applications, smaller than one.

The usual mathematical set-up for the treatment of induction and dynamo problems starts with the induction equation for the induced magnetic field

$$\frac{\partial \mathbf{b}}{\partial t} = \nabla \times (\mathbf{u} \times (\mathbf{B}_0 + \mathbf{b})) + \frac{1}{\mu_0 \sigma} \Delta \mathbf{b} \;, \qquad (3)$$

which follows from taking the *curl* of Amperes law $\nabla \times \mathbf{B} = \mu_0 \mathbf{j}$, using $\mathbf{j}$ given by Eq. (2), and inserting Faraday's law $\nabla \times \mathbf{E} = -\partial \mathbf{B}/\partial t$.

For some applications it is useful to switch over from the differential equation formulation to the integral equation formulation, which starts also from Eq. (2) and utilizes Biot-Savart's law for the magnetic field and Green's theorem for the electric potential [3,4]. In the following, we will only need the version for the steady case, in which the electric field can be expressed by the gradient of the electric potential $\mathbf{E} = -\nabla \phi$. The generalization to the time-dependent case can be found in [5].

The integral equation system consists of one equation for the induced magnetic field in the entire volume of the fluid,

$$\begin{aligned}\mathbf{b}(\mathbf{r}) = \;& \frac{\sigma \mu_0}{4\pi} \int_D \frac{(\mathbf{u}(\mathbf{r}') \times \mathbf{B}(\mathbf{r}')) \times (\mathbf{r} - \mathbf{r}')}{|\mathbf{r} - \mathbf{r}'|^3} dV' \\ & - \frac{\sigma \mu_0}{4\pi} \oiint_S \phi(\mathbf{s}') \, d\mathbf{S}' \times \frac{\mathbf{r} - \mathbf{s}'}{|\mathbf{r} - \mathbf{s}'|^3}\;,\end{aligned} \qquad (4)$$

and one equation for the electric potential at the fluid boundary:

$$\begin{aligned}\phi(\mathbf{s}) = \;& \frac{1}{2\pi} \int_D \frac{(\mathbf{u}(\mathbf{r}') \times \mathbf{B}(\mathbf{r}')) \cdot (\mathbf{s} - \mathbf{r}')}{|\mathbf{s} - \mathbf{r}'|^3} dV' \\ & - \frac{1}{2\pi} \oiint_S \phi(\mathbf{s}') \, d\mathbf{S}' \cdot \frac{\mathbf{s} - \mathbf{s}'}{|\mathbf{s} - \mathbf{s}'|^3}\;.\end{aligned} \qquad (5)$$

In Eq. (4) we see that the induced magnetic field $\mathbf{b}$ depends on both the induced primary currents in the volume $D$ and the induced electric potential $\phi$ at the fluid boundary $S$. Eqs. (4) and (5) are valid for arbitrary values of $Rm$. For small $Rm$, however, the induced magnetic field **b** under the volume integrals on the r.h.s. of Eqs. (4) and (5) can be neglected and the total **B** can be replaced by $\mathbf{B}_0$.

In [6] we had shown that the velocity structure of the flow can be reconstructed from the external measurement of an appropriate component of the induced magnetic field (e.g., the radial component for a spherically shaped fluid volume) and the induced electric potential at the fluid boundary. There remains a non-uniqueness concerning the radial distribution of the flow [7] what can be made plausible by representing the fluid velocity by two scalars living in the whole fluid volume (e.g., the toroidal and poloidal parts). Then it is clear that two quantities measured on a two-dimensional covering of the fluid give not enough information for the reconstruction of the two desired 3D-quantities.

In [8] we had developed a general method to avoid the electric potential measurement at the fluid boundary. The main idea is to apply the external magnetic field in two different, e.g. orthogonal, directions and to measure both corresponding sets of induced magnetic fields. The main obstacle for the application of this method comes from the electric potential term at the surface, the measurement of which we would like to circumvent. For the very special case of a spherical fluid volume, it can easily be seen, from multiplying Eq. (4) with the radial unit vector, that the radial component of **b** is unaffected by $\phi$. For other than spherical fluids, however, there is always an influence of the surface term on **b**. How can the influence of $\phi$ be taken into account without measuring it?

In the following we sketch the solution of this puzzle. Assume all measured induced magnetic field components be collected into an $NB$-dimensional vector with the entries $b_i$. Assume further a certain discretization of the electric potential at the surface, denoted by an $NP$-dimensional vector with the entries $\phi_m$. The velocity in the domain $D$ is discretized as an $NV$-dimensional vector with the entries $u_k$. Then, Eqs. (4) and (5) can be written in the form

$$b_i = K_{ik} u_k + L_{im} \phi_m \qquad (6)$$

And

$$\phi_m = M_{mk} u_k + N_{mn} \phi_n \qquad (7)$$

with **K** being a matrix of type $(NB, NV)$, **L** a matrix of type $(NB, NP)$, **M** a matrix of type $(NP, NV)$, and **N** a matrix of type $(NP, NP)$. Note that only the matrices **K** and **M** depend on the applied magnetic field $\mathbf{B}_0$, whereas the matrices **L** and **N** are independent of $\mathbf{B}_0$. Equation (7) can now be rewritten in the form

$$\phi_m = (I - N)^{-1, defl}_{mn} M_n u_n \qquad (8)$$

where we have used the superscript '*defl*' in order to indicate that the inversion of $\mathbf{I} - \mathbf{N}$ is a bit tricky due to its singularity, which reflects nothing but the physical irrelevance of the addition of a constant to $\phi$. '*defl*' is a shorthand for „deflation", a technique that has been used for long in the context of electro- and magnetoencephalography. Inserting Eq. (8) into Eq. (6), we get

$$b_i = K_{ik} u_k + L_{im} (I - N)^{-1, defl}_{mn} M_{nk} u_k \qquad (9)$$

which describes a linear connection between the vector **u** of all desired velocity components and the vector **b** of all magnetic field components measured for a given choice of $\mathbf{B}_0$.

**UNIQUENESS CONSIDERATIONS**

Having established, in Eq. (9), a linear relation between the desired velocity field and the measured magnetic fields, we can try to solve the inverse problem. The above mentioned non-uniqueness with regard to the depth dependence of the velocity can only be resolved by using applied magnetic fields with varying frequency.

Even then the problem arises whether the depth dependence of the velocity field can be recovered uniquely from a large enough set of frequency dependent data. In a first attempt to solve this problem we have considered a simplified dynamo problem in which the large-scale velocity is replaced the helical turbulence parameter $\alpha$, which is an important quantity in dynamo theory. Choosing this parameter spherically symmetric and isotropic, we end up with the inverse spectral problem to determine the radial dependence of $\alpha$ from a given set of eigenvalues of a matrix eigenvalue equation for the defining scalars of the magnetic field. This inverse spectral problem has been tackled both numerically [9] and analytically [10]. Still, the results are not completely conclusive, but it is clear that the use of data with varying frequencies could mitigate the non-uniqueness of the inversion with respect to the depth dependence of the velocity. A nice byproduct of the numerical work, which uses an evolution strategy, was the construction of truly oscillatory $\alpha^2$ dynamos with spherically symmetric $\alpha$, whose existence had been questioned for a long time [11].

## REGULARIZATION ISSUES

For the data analysis of the experiment we circumvent the non-uniqueness problem by the use of Tikhonov regularization and the method of Tikhonov's L-curve [12]. Note that the inversion of Eq. (9) is equivalent to the minimization of the mean quadratic deviation of the measured fields $b^{meas}_{i,B_{01/2}}$ from the computed fields $b^{comp}_{i,B_{01/2}}[\mathbf{v}]$ resulting from a given velocity distribution $\mathbf{v}$. Within the regularization one adds to the functional of mean quadratic residual one or more „penalty" functions which are chosen, for convenience, as quadratic functionals of the velocity. For our purpose we minimize the total cost functional

$$F[\mathbf{v}] = F_{B_{01}}[\mathbf{v}] + F_{B_{01}}[\mathbf{v}] + F_{div}[\mathbf{v}] + F_{pen}[\mathbf{v}] \quad (10)$$

with the individual parts

$$F_{B_{01}}[\mathbf{v}] = \sum_{i=1}^{NB} \frac{1}{\sigma^2_{i,B_{01}}} \left( b^{meas}_{i,B_{01}} - b^{comp}_{i,B_{01}}[\mathbf{v}] \right)^2 \quad (11)$$

$$F_{B_{02}}[\mathbf{v}] = \sum_{i=1}^{NB} \frac{1}{\sigma^2_{i,B_{02}}} \left( b^{meas}_{i,B_{02}} - b^{comp}_{i,B_{02}}[\mathbf{v}] \right)^2 \quad (12)$$

$$F_{div}[\mathbf{v}] = \frac{1}{\sigma^2_{div}} \sum_{k=1}^{NV} (\nabla \cdot \mathbf{v})^2_k \Delta V_k \quad (13)$$

$$F_{pen}[\mathbf{v}] = \frac{1}{\sigma^2_{pen}} \sum_{k=1}^{NV} \mathbf{v}^2_k \Delta V_k \quad (14)$$

which represent the mean quadratic residuals of the two data sets measured for $\mathbf{B}_{01}$ and $\mathbf{B}_{02}$, the functional that minimizes the divergence of the velocity field, and the quadratic penalty function, respectively. The minimization of the total cost functional in Eq. (10) is done by standard techniques. By scaling the penalty parameter $\sigma_{pen}$ one can derive the so-called Tikhonov's L curve, i.e., the mean squared residual in dependence on the penalty function. At the point of highest bending (the „knee") we get a good compromise between the fit of the model to the measured data on one hand and the penalty function on the other hand [12].

## EXPERIMENTAL SET-UP

In order to demonstrate the feasibility of the contactless inductive velocity reconstruction method in real applications an experiment has been set up in which the propeller driven flow of a liquid metal has to be reconstructed solely from externally measured magnetic field data.

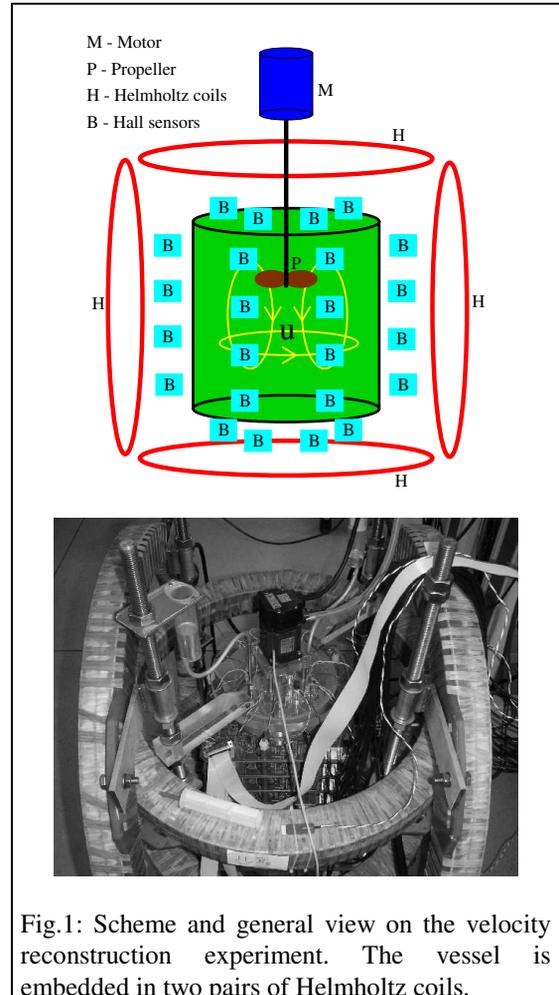

Fig.1: Scheme and general view on the velocity reconstruction experiment. The vessel is embedded in two pairs of Helmholtz coils.

Figure 1 shows the experimental set-up. We use 4.4 liters of the eutectic alloy $Ga^{67}In^{20.5}Sn^{12.5}$ which is liquid at room temperatures. Its density and electrical conductivity at 20°C are $6.36 \times 10^3$ kg/m$^3$ and $3.31 \times 10^6$ S/m, respectively. The flow is produced by a motor driven propeller with a diameter of 6 cm inside a polypropylene vessel with 18.0 cm diameter. The height of the liquid metal is 17.2 cm, giving an aspect ratio close to 1.

The propeller can rotate in both directions, resulting either in upward or downward pumping. The rotation rate can reach 2000 rpm producing a mean velocity of 1 m/s, which corresponds to an $Rm \approx 0.4$. The flow structure for the two directions is not symmetric for three reasons. At first, the propeller itself has a preferred direction. At second, it is positioned approximately at one third of the liquid height, measured from the top. At third, there are 8 guiding blades above the propeller, having two functions. For the downward pumping, they are intended to remove the upstream swirl in order not to deteriorate the pressure conditions for the propeller. For the upward pumping they remove the downstream swirl thus leaving the entire flow rather swirl-free.

Hence, the downward pumping produces, in addition to the main poloidal roll, a considerable toroidal motion, too. For the upward pumping, this toroidal motion is, to a large extend, inhibited by the guiding blades. It was one of the tasks of the experiment to discriminate between those flow structures.

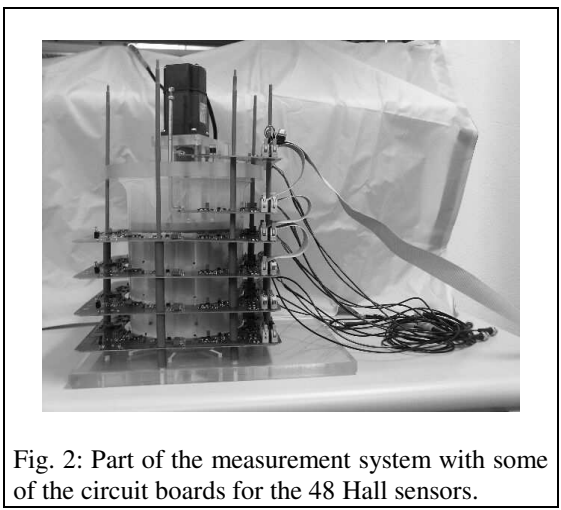

Fig. 2: Part of the measurement system with some of the circuit boards for the 48 Hall sensors.

The axial and the transversal magnetic fields are produced by two pairs of Helmholtz coils. With a current of 55 A they produce a field of 8 mT that is rather homogeneous and isotropic throughout the vessel.

The induced magnetic fields are measured with 48 Hall sensors, 8 of them grouped together on each of the 6 circuit boards which are located on different heights (Figs. 1, 2). One additional sensor is located in the center below the vessel.

The basic problem for any practical application of the method is the reliable determination of small induced magnetic fields on the background of much larger imposed magnetic fields. For this it is indispensable to have an accurate control of the external magnetic field. In our configuration the current drift in the Helmholtz coils can be controlled with an accuracy of better than 0.1 per cent. This is sufficient since the measured induced fields are approximately 1 per cent of the applied field.

Another serious problem is the unavoidable temperature drift of the Hall sensors. This concerns their sensitivity as well as the offset. The sensitivity drift is connected with the temperature dependence of the semiconductors resistance and can be overcome by keeping the current constant. The offset problem is overcome by changing the sign of the applied magnetic field. For that purpose, we choose a trapezoidal signal form which changes its sign every 0.5-1 sec. After the measurement has been accomplished for the applied transversal field $B_{0x}$, the same procedure is carried out for the applied axial field $B_{0z}$.

**EXPERIMENTAL RESULTS**

For both directions of the applied magnetic field, the induced magnetic fields are measured and put into the inverse problem solver.

Figures 3 and 4 show the induced magnetic fields for the cases of upward and downward pumping, respectively. In each figure, the upper part shows the induced field for the applied transversal field $B_{0x}$, and the lower part shows the induced magnetic field for the applied axial field $B_{0z}$.

For both flow directions we show in Fig. 5 the Tikhonov's L-curves with a clear-cut „knee" which is interpreted as the point with the physically most reasonable solution. The penalty parameter at this point, which is approximately the same for both flow directions, can be used in the online monitoring of the flow. The two plotted curves correspond to the inversion for the cases that we truly consider in Eq. (4) the surface term, and that we neglect this term.

Figures 6 and 7 show the results of the inversion. In Fig. 6, we see clearly the upward flow in the center of the vessel and the downward flow at the rim, but nearly no rotation of the flow. In Fig. 7 we see the downward flow together with a clear rotation of the flow. This absence and presence of the swirl is an important feature which can evidently be discriminated by our method.

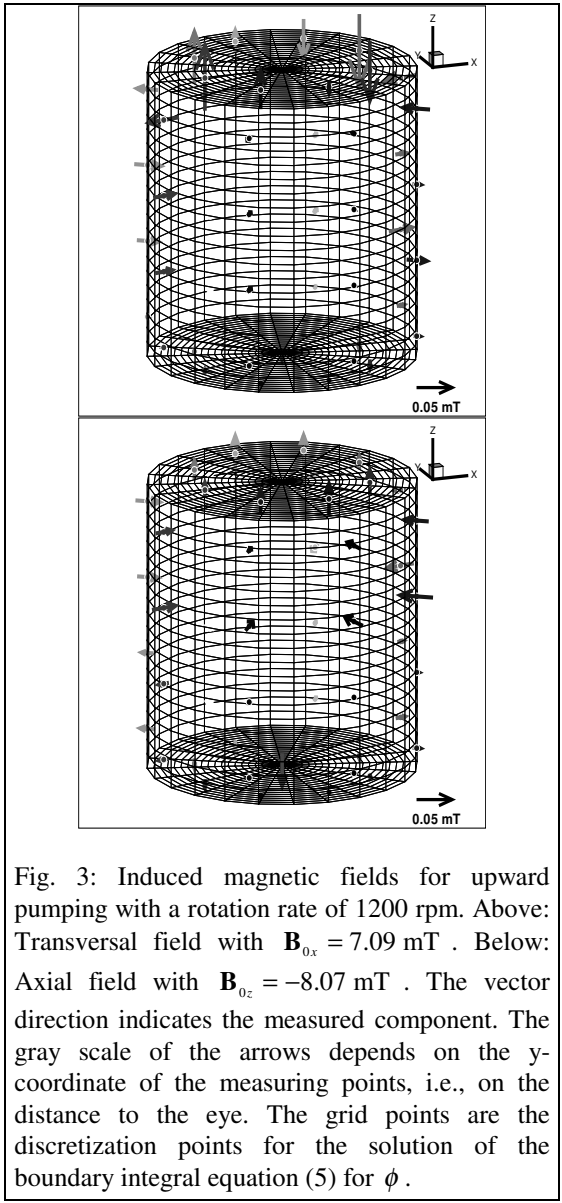

Fig. 3: Induced magnetic fields for upward pumping with a rotation rate of 1200 rpm. Above: Transversal field with $\mathbf{B}_{0x} = 7.09$ mT. Below: Axial field with $\mathbf{B}_{0z} = -8.07$ mT. The vector direction indicates the measured component. The gray scale of the arrows depends on the y-coordinate of the measuring points, i.e., on the distance to the eye. The grid points are the discretization points for the solution of the boundary integral equation (5) for $\phi$.

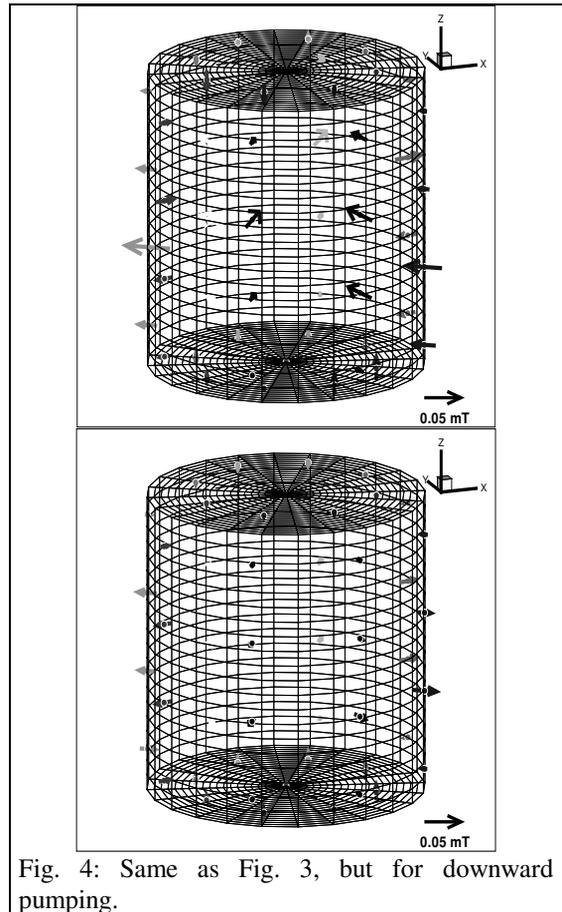

Fig. 4: Same as Fig. 3, but for downward pumping.

It is worth to note that not only the structure of the flow, but also the range of the velocity scale is correctly reproduced by the inversion.

We have restricted our inversion to the determination of the velocity at only 45 points. Considering only two components at every point as independent (the third can be fixed by the divergence-free constraint) we end up with 90 degrees of freedom which are to be determined from $2 \times 49 = 96$ measured data. Having more information than wanted quantities, we could get a reasonable result even without regularization. This is seen by the fact that in Fig. 5 the Tikhonov's L-curves are stuck at the r.h.s. without decreasing down to a zero mean squared residual. This decrease would happen for a larger number of wanted velocity degrees of freedom.

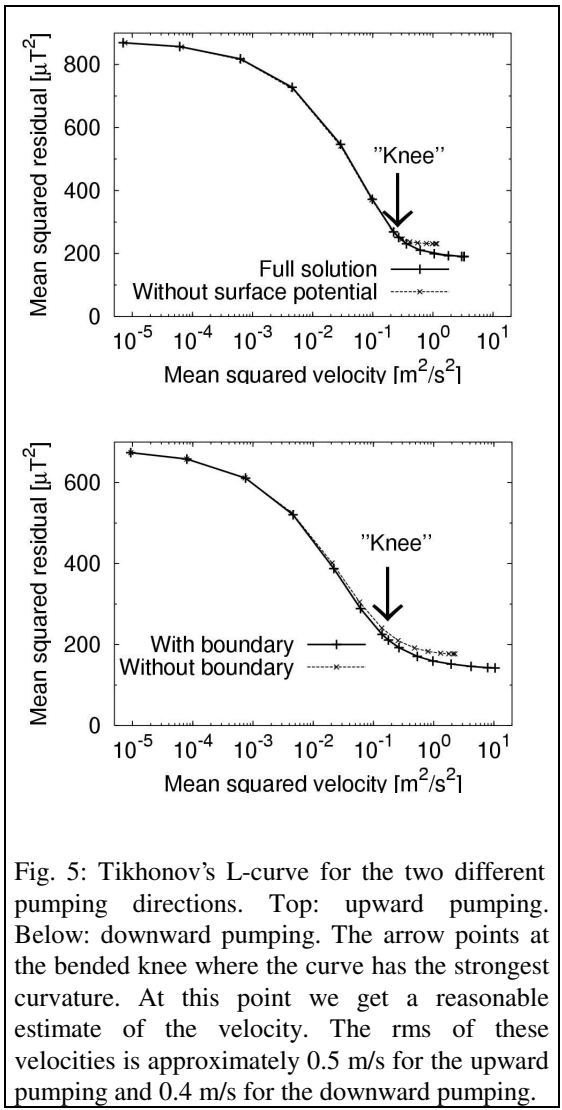

Fig. 5: Tikhonov's L-curve for the two different pumping directions. Top: upward pumping. Below: downward pumping. The arrow points at the bended knee where the curve has the strongest curvature. At this point we get a reasonable estimate of the velocity. The rms of these velocities is approximately 0.5 m/s for the upward pumping and 0.4 m/s for the downward pumping.

Even at the present level of data acquisition, without imposing magnetic fields with varying frequencies, there is room for improvements of the inversion scheme. For example, one could represent from the very beginning the velocity components by a finite number of smooth functions and determine the expansion coefficients in the inversion. At least, this would significantly improve the appearances of the resulting velocity field, if compared to the rough point-wise vector representation used in this paper. However, even with apparently smooth velocity fields one should not forget that the number of reliable degrees of freedom cannot be larger than the number of the measured degrees of freedom.

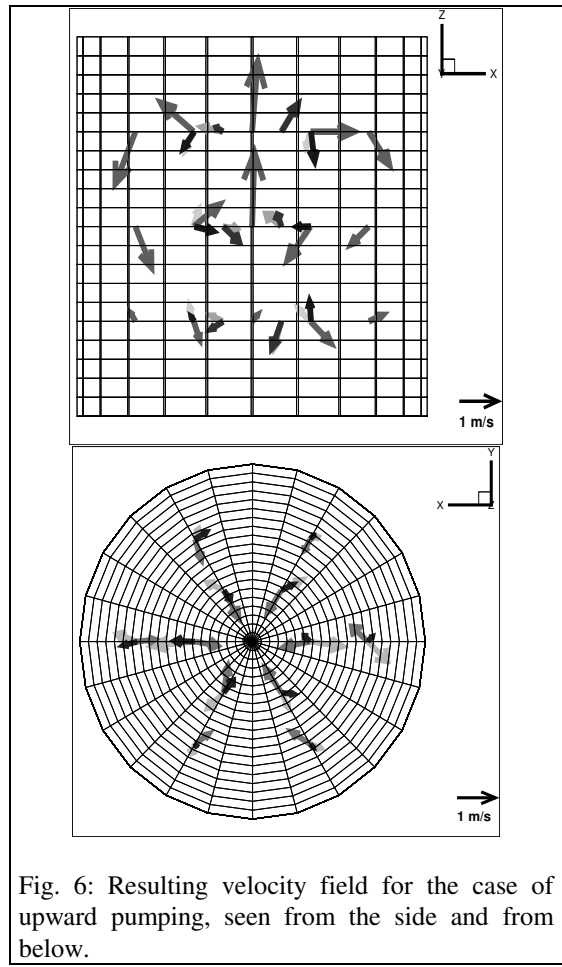

Fig. 6: Resulting velocity field for the case of upward pumping, seen from the side and from below.

## Conclusions and Prospects

Flow measurement based on induced electric potentials has a long tradition and a lot of practical applications [13]. The change from induced electric potentials to induced magnetic fields allows a completely contactless measuring technique for those flows whose magnetic Reynolds number is not too small. In a liquid metal experiment, we have put into practice the theory of 3D velocity reconstruction from externally measured magnetic fields that had been developed by us during the last years. The challenges for the future lay with improving the depth resolution of the method by using magnetic fields with variable frequency on one hand, and with the extension of the method to inverse dynamo problems with larger magnetic Reynolds numbers, on the other hand.

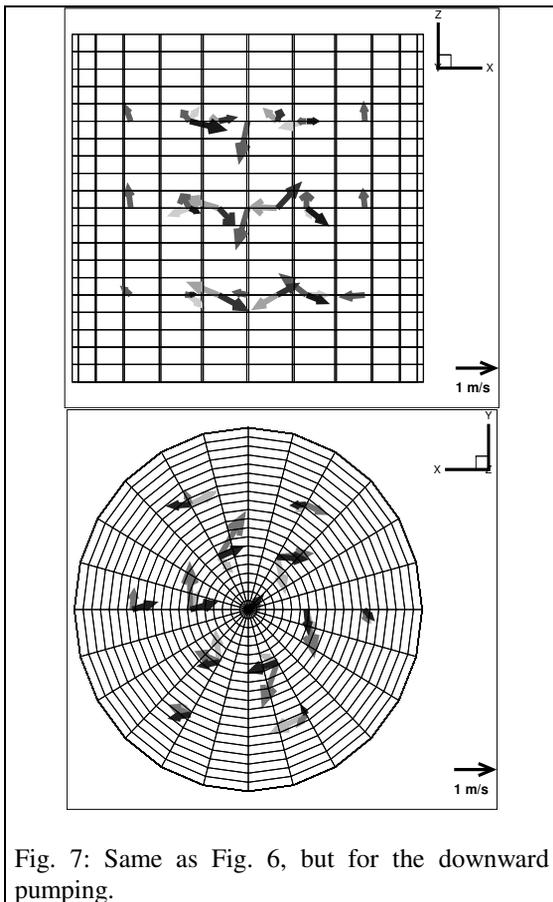

Fig. 7: Same as Fig. 6, but for the downward pumping.